\documentclass[aps,prd,showpacs,twocolumn,superscriptaddress,
preprintnumbers,10pt]{revtex4}
\usepackage{amssymb}
\usepackage{graphicx}
 
\begin{document}
 
\title{The High-Density Symmetry Energy and Direct Urca}
\author{Andrew W. Steiner}
\affiliation{Theoretical Division, Los Alamos National Laboratory,
Los Alamos, NM 87545, USA}
 
\preprint{LA-UR-06-4517}
 
\begin{abstract}
The symmetry energy of nucleonic matter is usually assumed to be
quadratic in the isospin density. While this may be justified at
sub-saturation densities, there is no need to enforce this restriction
at super-saturation densities. The presence of a quartic term can
strongly modify the critical density for the direct Urca process which
leads to faster cooling of neutron stars. Neutron star cooling
predictions which lie below the observational data can, for some
equations of state, be repaired with a quartic term which effectively
turns off the direct Urca process.
\end{abstract}
 
\pacs{97.60.Jd, 26.60.+c, 21.65.+f}

\maketitle

\section{Introduction}

The equation of state of nucleonic matter is often decomposed into an
isospin-symmetric and an isospin asymmetric part
\begin{eqnarray}
E(n,\delta) = E_{\mathrm{nuc}}(n,\delta=0) + \delta^2
S(n) + \delta^4 Q(n) + ... \, ,
\label{eq:qex}
\end{eqnarray}
where $n=n_n+n_p$, $\delta=1-2 n_p/n$, $E_{\mathrm{nuc}}$ is the
energy per baryon of nuclear matter and $S$ and $Q$ are arbitrary
functions of density. Most often, $Q$ is sufficiently small so that it
can be ignored, and then $E_{\mathrm{sym}}=S(n)$ is the
density-dependent and isospin-independent symmetry energy.

Without the Coulomb interaction, the near isospin symmetry of QCD
predicts that pure neutron matter and pure proton matter should have
the same energy density.  Terms in Eq.~\ref{eq:qex} with odd-powers in
delta are thus forbidden, assuming the $\delta$ expansion is
analytic. Near the nuclear saturation density, it has been suggested
that the truncation to order $\delta^2$ for the potential energy part
of the equation of state is a very good approximation for densities
below or near the nuclear saturation density, 0.16
fm$^{-3}$~\cite{Lagaris81,Prakash87,Wiringa88,Muther88,Bombaci91,Lee98}. In
this article, the effect of the quartic term in Eq.~\ref{eq:qex} at
higher density is examined, where no constraints on the quartic terms
are known because the uncertainty contained in many-body effects.

The major effect of a ``quartic term'' (i.e. a term that is of
fourth-order in the isospin deviation, $\delta$) is to change the
ratio of protons to neutrons at higher densities. All other things
remaining equal, the effect on neutron star masses and radii from a
quartic term is typically small. However, the critical density for the
direct Urca cooling process~\cite{Lattimer91},
\begin{eqnarray}
n \rightarrow p + e + \bar{\nu}_e 
\quad\mathrm{and}\quad
p + e \rightarrow n + \nu_e \, ,
\end{eqnarray}
is very sensitive to the number of protons present in dense
matter. This process can only proceed when the conditions for momentum
and energy conservation can be simultaneously satisfied, when the
proton fraction is larger than about 10 percent. The presence
of a quartic term in Eq.~\ref{eq:qex} can either move the critical
density for the direct Urca process to much lower densities, or to
densities larger than the central density of the maximum mass neutron
star. 

Neutron stars cool much more quickly when the direct Urca process is
allowed~\cite{Page92,Lattimer94}. Thus understanding the possible 
magnitude of quartic terms is essential in describing the cooling of
neutron stars.

\section{Origins of Quartic Terms}

For the present discussion, it is useful to separate the equation of
state into contributions to the potential and kinetic energy. While
this demarcation is not necessarily unique, we use the term ``kinetic
energy part'' to refer to the Fermi gas portion (including the
interactions present in the nucleon effective mass) of the equation of
state (whether relativistic or non-relativistic) and ``potential
energy part'' to refer to the remainder. The kinetic energy part of
the symmetry energy naturally contains quartic terms in all equations
of state. The potential energy part of the equation of state is the
primary consideration here, but it should be noted that interactions
can affect the size of quartic terms in the kinetic energy through
their effect on the nucleon effective masses.

In relativistic mean-field
models~\cite{Walecka74,Boguta77,Serot79,Muller96,Furnstahl02} the
many-body interactions have a non-trivial isospin dependence even in
their most trivial form. The linear Walecka model, for example,
already has quartic terms (albeit small) in the potential part of the
energy density. These quartic terms can be much stronger when three-
and higher-body contributions are included. An example is a term
proportional to $\sigma \rho^2$ where $\sigma$ is the scalar-isoscalar
meson and $\rho$ is the vector-isovector meson.  In the context of a
relativistic point-coupling
model~\cite{Rusnak97,Nikolaus97,Burvenich02}, such a term would
correspond to a three-body nucleon-nucleon interaction of the form
\begin{eqnarray}
\bar{\psi} \psi (\bar{\psi} \gamma^{\mu} \vec{\tau} \psi) \cdot 
(\bar{\psi} \gamma_{\mu} \vec{\tau} \psi)
\end{eqnarray}

In Skyrme-like models, no quartic term is present in the potential
energy part of the equation of state. The two-body interaction present
in the Skyrme model results in two powers of density, and the
three-body interaction is treated as a density-dependent two-body
interaction where the additional density dependence is assumed to be
exactly isospin symmetric.

Modern microscopic-macroscopic~\cite{Moller95} and
Hartree-Fock-based~\cite{Samyn02} models of nuclear masses contain
explicit non-quadratic terms in the form of the phenomenological term
called the Wigner energy. This term is often thought to originate in
isovector
pairing~\cite{Satula97,Satula97b,Satula98,Vogel00,Macchiavelli00,Afanasjev05}. Pairing
is thought to be primarily a surface effect and vanishes in the
infinite baryon number limit. Thus it does not contribute to the
equation of state of infinite nucleonic matter here. However, if the
Wigner energy was shown to have a different origin, or if the pairing
energy was shown to be partially a volume effect (rather than a
surface effect), then it would contribute to the isospin dependence of
the symmetry energy. Neutron star matter, however, is typically
sufficiently neutron-rich that isovector pairing is not a strong
contribution.

\section{Convenient Parameterization of the Quartic Dependence}

A standard alternative to Eq.~\ref{eq:qex} is to define the 
symmetry energy using the second derivative with respect to the 
isospin asymmetry
\begin{eqnarray}
E_{\mathrm{sym}}(n,\delta) = \frac{1}{2}
\frac{d^2 E(n,\delta)}{d \delta^2} = S(n)+6 Q(n) \delta^2
\, ,
\label{eq:ddef}
\end{eqnarray}
where $\varepsilon$ is the energy density.  In this work,
Eq.~\ref{eq:ddef} will be taken to be the definition of the symmetry
energy.  If there are no quartic terms in Eq.~\ref{eq:qex}, then
$E_\mathrm{sym}=S(n)$ is independent of the isospin asymmetry,
$\delta$.  In the presence of quartic terms, this definition of the
symmetry energy in Eq.~\ref{eq:ddef} is more closely connected to the
experimental observables based on nuclei which like close to the $N=Z$
line than the alternative of defining the symmetry energy as the
energy difference between pure neutron matter and isospin-symmetric
nuclear matter.

A convenient way to parameterize the addition of a quartic term
is to define $\eta$ as
\begin{eqnarray}
\eta(n) = \frac{E(n,1)-E(n,1/2)}{3 [E(n,1/2)-E(n,0)]} 
= \frac{4 S(n) + 5 Q(n)}{4 S(n) + Q(n)}
\, .
\label{eq:hdef}
\end{eqnarray}
This parameter is unity when the quartic term is zero and greater or
less than one depending on whether the symmetry energy is increased or
decreased by the quartic term. This parameter is more convenient for
numerical work than using the generalization of
Eq.~\ref{eq:ddef}~\cite{Lee98},
\begin{eqnarray}
E_{\mathrm{sym},4}(n) = \frac{1}{24}
\frac{d^4 E(n,\delta)}{d \delta^4} = Q(n) \, ,
\label{eq:s4}
\end{eqnarray}
since numerical fourth derivatives can be difficult to compute
accurately. The value of $\eta$ is restricted by ensuring that the
derivative of the energy per baryon as a function of delta does not
vanish at any point other than at $\delta=0$ corresponding to
isospin-symmetric nuclear matter. This guarantees that, at fixed
density, nuclear matter is always the most energetically favored
configuration and neutron matter is always the most costly
configuration (when ignoring the Coulomb interaction and possible
constraints from $\beta$-equilibrium). The proper limits on $\eta$,
$3/7 < \eta < 5$ are easily obtained from the definition
above. Sometimes it is useful to distinguish the value of $\eta$ as
obtained from the ``kinetic'' part of the EOS, $\eta_{\mathrm{kin}}$,
that obtained from the ``potential'' part of the EOS,
$\eta_{\mathrm{pot}}$, and that obtained from the full EOS,
$\eta_{\mathrm{tot}}$. Note that $\eta_{\mathrm{tot}}$ is not
trivially related to the individual contributions
$\eta_{\mathrm{kin}}$ and $\eta_{\mathrm{pot}}$. 

Enforcing a particular density-dependence of $\eta$ on the potential
part of a given equation of state is straightforward. Defining ${\cal
T}$ as
\begin{eqnarray}
{\cal T}(x_p) &=& \frac{4}{1+3 \eta}\left[
-3 x_p +19 x_p^2 -32 x_p^3+16 x_p^4 + \right.
\nonumber \\ && 
\left. 7 x_p \eta -23 x_p^2 \eta + 32 x_p^3 \eta-16 x_p^4 \eta
\right] \, ,
\label{eq:calt}
\end{eqnarray}
where $x_p=n_p/n$, one can construct a new equation of state from any
equation of state of neutron and nuclear matter using
\begin{eqnarray}
E_{\mathrm{pot}} (n,\delta) 
= {\cal T} E_{\mathrm{pot,nuc}}(n) + (1-{\cal T}) E_{\mathrm{pot,neut}}(n)
\, .
\end{eqnarray}
(Note that the limit $\eta \rightarrow 1$ gives the correct expression
for a purely quadratic symmetry energy, $1-(1-2 x_p)^2$.) Only the
potential energy part of the equation of state has been modified, and
the kinetic energy part remains unchanged.  The quantity ${\cal T}$ is
density dependent if $\eta$ is density dependent, as is typical for
many equations of state. The corresponding potential parts of the
neutron and proton chemical potentials are given by
\begin{eqnarray}
\mu_{\mathrm{pot},n} &=& {\cal T} \mu_{\mathrm{pot,nuc},n} + 
(1-{\cal T}) \mu_{\mathrm{pot,neut},n} +
\nonumber \\
&& T^{\prime}(x_p) \frac{(x_p-1)}{n} \left( \mu_{\mathrm{pot,neut},n} -
\mu_{\mathrm{pot,nuc},n} \right)
\nonumber \\
\mu_{\mathrm{pot},p} &=& {\cal T} \mu_{\mathrm{pot,nuc},p} + 
(1-{\cal T}) \mu_{\mathrm{pot,neut},p} + 
\nonumber \\
&& T^{\prime}(x_p) \frac{x_p}{n} \left( \mu_{\mathrm{pot,neut},p} -
\mu_{\mathrm{pot,nuc},p} \right)
\label{eq:new_mu}
\end{eqnarray}
The effective masses are not modified here since they are in the
kinetic energy part of the EOS. 

\begin{figure}[htb]
\includegraphics[scale=0.37]{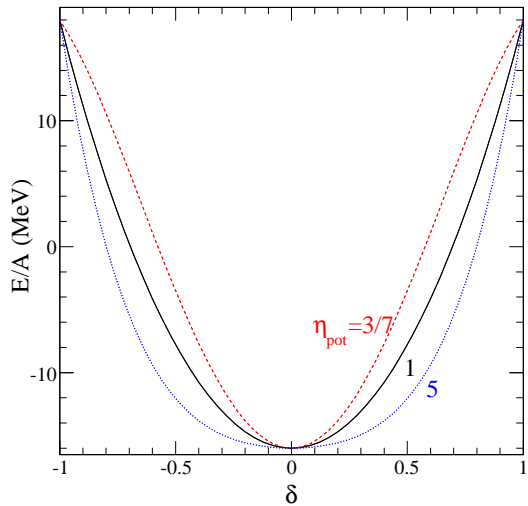}
\caption{The energy per baryon of the APR EOS at saturation density
as a function of the proton fraction, $x_p$, for the trivial and
limiting values of $\eta_{\mathrm{pot}}$.}
\label{fig:apr_demo}
\end{figure}

The behavior of the energy per baryon as a function of $\delta$ at
saturation density in the equation of state (EOS) of Akmal, et. al
(APR).~\cite{Akmal98} constructed as above to have varying values of
$\eta_{\mathrm{pot}}$ is given in Fig.~\ref{fig:apr_demo}. The
deviation is not large, at most a couple MeV near saturation density for
the most extreme values of $\eta$.

\section{The Value of $\eta$ for Modern Equations of State}

In addition to APR from above, several relativistic mean-field (RMF)
models are used in this work, all of which are constrained to have the
following properties at saturation
\begin{eqnarray}
\mathrm{n}_0 &=& 0.16~\mathrm{fm}^{-3} \nonumber \\
\mathrm{E/A} &=& -16~\mathrm{MeV} \nonumber \\
\mathrm{K}&=&230~\mathrm{MeV} \nonumber \\
\mathrm{E}_{\mathrm{sym}}&=&34~\mathrm{MeV} \nonumber \\
\mathrm{M^{*}/M} &=& 0.8 
\label{eq:rmf_sat}
\end{eqnarray}
where $\mathrm{n}_0$ is the saturation density, $\mathrm{E/A}$ is the
binding energy, $\mathrm{K}$ is the compressibility,
$\mathrm{E}_{\mathrm{sym}}$ is the symmetry energy at the saturation
density, and $\mathrm{M^{*}/M}$ is the reduced Dirac effective mass. In
addition to the standard RMF model with no non-linear couplings except
for the cubic and quartic self-interactions among the scalar mesons,
two RMF models are constructed with small (``RMFlo'') and large
(``RMFhi'') values of $\eta_{\mathrm{pot}}$ in order to demonstrate
the variations that are possible within this formalism. These two
models do not necessarily represent the most extreme cases, as a full
optimization was not performed. The couplings are given in Table 1 and
utilize the Lagrangian and notation described in~\cite{Steiner05}. All
of these couplings are within a factor of two of the constraints
required by naturalness~\cite{Friar96,Muller96}.

\begin{table}
\caption{
\label{tab:table1}
The couplings for the RMF, RMFlo, and RMFhi models. The units are all
given so that the Lagrangian has units $\mathrm{MeV}^4$ when the
meson fields are in $\mathrm{MeV}$ and the nucleon fields in
$\mathrm{MeV}^{3/2}$. Omitted couplings are equal to zero.
}
\begin{ruledtabular}
\begin{tabular}{cccc}
RMF & & & \\
\hline
$g_{\sigma}$ & 7.721 & $g_{\omega}$ & 7.955 \\
$g_{\rho}$ & 8.608 & $\kappa$ & 21.40 \\
$\lambda$ & -6.227 $\times 10^{-4}$ & & \\
\hline
RMFlo & & & \\
\hline
$g_{\sigma}$ & 7.692 & $g_{\omega}$ & 7.998 \\
$g_{\rho}$ & 14.00 & $\kappa$ & 18.9 \\
$\lambda$ & 5.188 $\times 10^{-3}$ & $\zeta$ & 0.03801 \\
$\xi$ & 1.499 & $a_1$ & 69.6 \\
$a_2$ & 1.052 & $b_1$ & 0.05251 \\
$a_3$ & 8.446 $\times 10^{-3}$ & $a_4$ & -2.063 $\times 10^{-5}$ \\
$b_2$ & -8.085 $\times 10^{-6}$ & & \\
\hline
RMFhi & & & \\
\hline
$g_{\sigma}$ & 7.677 & $g_{\omega}$ & 8.021 \\
$g_{\rho}$ & 7.551 & $\kappa$ & 17.50 \\
$\lambda$ & 8.361 $\times 10^{-3}$ & $\zeta$ & 0.05880 \\
$\xi$ & 0.02870 & $a_1$ & -35.30 \\
$a_2$ & -0.66605 & $b_1$ & 0.2719 \\
$a_3$ & 1.281 $\times 10^{-3}$ & $a_4$ & 2.062 $\times 10^{-5}$ \\
$b_2$ & 1.127 $\times 10^{-5}$ & & \\
\hline
\end{tabular}
\end{ruledtabular}
\end{table}
The Skyrme~\cite{Skyrme59} model ``SLy230a''~\cite{Chabanat97} is used
for comparison. This model matches the binding energies and charge
radii of several ground-state nuclei, the known properties of
saturated nuclear matter, and results in reasonable neutron star
properties.

The momentum dependence of the nucleon optical potential is essential
for transport simulations of heavy-ion collisions. The MDI
(momentum-dependent interaction) EOS~\cite{Das03}, which has
successfully described aspects of heavy-ion collisions at intermediate
energies, is included here. In addition to providing for reasonable
properties of saturated nuclear matter, the isoscalar potential of the
MDI EOS coincides with predictions from the variational many-body
theory using inputs constrained by nucleon-nucleon scattering
data~\cite{Wiringa88,Li04a}, and the isovector potential agrees with
the momentum dependence of the Lane potential extracted from (p,n)
charge exchange reactions up to about 45 MeV~\cite{Hoffmann72,Li04a}.
We use the formulation as described in Ref.~\cite{Das03} with the symmetry
energy parameter, $x$, set equal to zero.

The APR and Skyrme EOSs are special in that $\eta_{\mathrm{pot}}=1$
independent of density. In the case of the Skyrme models, this results
from the structure of the interaction, and in the case of APR,
$\eta_{\mathrm{pot}}=1$ by construction.

\begin{figure}[htb]
\includegraphics[scale=0.37]{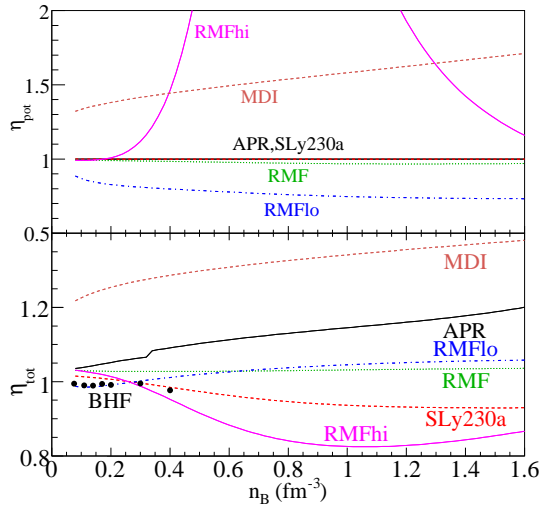}
\caption{The density dependence of $\eta_{\mathrm{pot}}$ and
$\eta_{\mathrm{tot}}$ determined from Eq.~\ref{eq:hdef}
for the equations of state considered in this
work as described in the text. The kink in APR is due to the phase
transition to the high-density phase.}
\label{fig:aden}
\end{figure}

In Fig.~\ref{fig:aden}, the intrinsic values of $\eta$ (the procedure
described in Eqs.~\ref{eq:calt}-\ref{eq:new_mu} has not yet been
employed) as obtained from the EOSs described above are plotted, as
well as that obtained by fitting the results in
Ref.~\cite{Bombaci91}. Note that only the total energy per baryon was
reported in this latter reference, so there are no results for the top
panel. The value of $\eta_{\mathrm{pot}}$ for the typical RMF model is
quite close to unity, while RMFlo and RMFhi (by construction) have
values significantly different from 1. The maximum value of
$\eta_{\mathrm{pot}}$ for the RMFhi EOS, even though it lies outside
the graph, is below the limit $\eta<5$ mentioned in the
introduction. The MDI EOS has a large value of $\eta_{\mathrm{pot}}$,
even at lower densities where most models have
$\eta_{\mathrm{pot}}~\sim 1$

The values of $\eta_{\mathrm{tot}}$ shown in the bottom panel show a
little more variation. The value of $\eta_{\mathrm{tot}}$ from the BHF
calculation is quite close to 1, even at lower densities where other
models (except RMFlo) predict larger values. This is typical; in order
to get $\eta_{\mathrm{tot}}$ equal to unity at low densities, one
frequently has $\eta_{\mathrm{pot}}<1$ in order to cancel the effect
of $\eta_{\mathrm{kin}}$ which is most often larger than one. It is
also quite clear that the effects of the kinetic part of the equation
of state are not trivial in determining $\eta_{\mathrm{tot}}$. The
ordering of the values of $\eta$ for the RMFlo and RMFhi EOSs reverses
depending on whether or not one is considering the total equation of
state or only the potential energy part.

\section{Direct URCA}

When the neutron star temperature is small enough that it can be
ignored in comparison to the Fermi momenta of the consituents, the
critical density for direct Urca is equivalent to the condition that
one be able to form a triangle with the neutron, proton, and electron
Fermi momenta. This triangle may be formed if the squared area
obtained by Heron's formula is positive, i.e.
\begin{equation}
s(s-k_{F n})(s-k_{F p})(s-k_{F e}) > 0
\end{equation}
where $s$ is the semiperimeter defined by $s = (k_{F n}+k_{F p}+k_{F
 e})/2$. This reduces the familiar condition $k_{F n} < k_{F p}+k_{F
 e}$ for neutron-rich matter and is more easily generalizable to 
other Urca-like processes.

\begin{figure}[htb]
\includegraphics[scale=0.37]{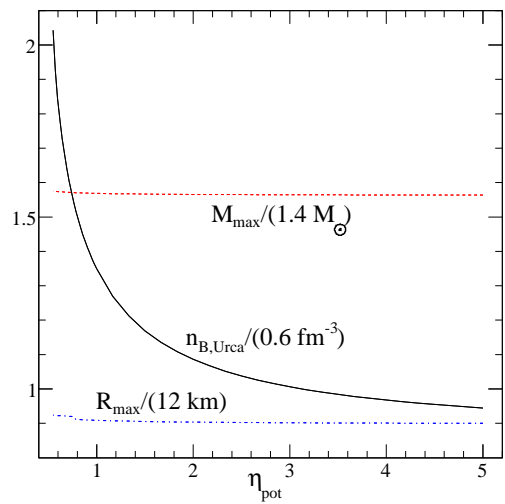}
\caption{The critical density for the direct Urca process for the APR
EOS as a function of $\eta_{\mathrm{pot}}$. Points with
$\eta_{\mathrm{pot}}<1/2$ were not plotted because the Urca process is
not allowed at any density for this range of $\eta$.}
\label{fig:apr_urca}
\end{figure}

When an EOS has a proton fraction which is small, true of the APR EOS
and some EOSs based on Skyrme interactions, the presence of a quartic
term can drastically affect the critical density for the direct Urca
process. This is demonstrated in Fig.~\ref{fig:apr_urca}, where the
critical density is plotted as a function of $\eta_{\mathrm{pot}}$,
assuming that $\eta_{\mathrm{pot}}$ is density independent. Depending
on the relative strength of the quartic dependence, Urca either
proceeds in all neutron stars with masses larger than about 1.4 solar
masses or in no neutron stars of any mass. Although values of
$\eta_{\mathrm{pot}}$ less than 1/2 are allowed, the corresponding
critical densities are already above the central density for the
maximum mass configuration. The maximum mass and radius of the maximum
mass neutron star as a function of $\eta_{\mathrm{pot}}$ are also
plotted in Fig.~\ref{fig:apr_urca}. These quantities are essentially
unaffected by the modification of $\eta_{\mathrm{pot}}$.

RMF models typically have a large symmetry energy and a large proton
fraction, and thus allow the direct Urca process at low
densities. However, this is not required, and Ref.~\cite{Steiner05}
constructed a couple RMF models with small neutron star radii (named
SR1, SR2, and SR3) which do not allow direct Urca at any density lower
than the central density of the maximum mass configuration.  These
models also typically have a smaller symmetry energy at saturation
density than the models utilized here. The models RMF,
RMFlo, and RMFhi, have critical densities of $0.306$, $0.275$, and
$0.806$ fm$^{-3}$, respectively. Note that RMFhi has a rather large
critical density for direct Urca which could easily be made larger by
decreasing the value of the symmetry energy at saturation
density. 

Ref.~\cite{Li05} compared the radii of 1.4 $\mathrm{M}_{\odot}$ neutron
stars with the MDI EOS using different symmetry energies and comparing
with isospin diffusion data from intermediate-energy heavy-ion
collisions. This reference found that the threshold for the direct
URCA process was signficantly different for the APR EOS and for the
MDI EOS (with $x=0$) even though they had nearly identical symmetry
energies. This is, in large part, due to the signficant presence of
quartic terms in the MDI EOS as demonstrated above. This
does not contradict the constraint on neutron star radii from
Ref.~\cite{Li05}, since neutron star radii are insensitive to quartic
terms, as demonstrated in Fig.~\ref{fig:apr_urca}.

\section{Discussion}

Quartic terms play an important role in determining the critical
density for the direct Urca process in neutron stars. These terms can
be easily generated within the context of RMF models of high-density
nucleonic matter. While this work means that it will be more difficult
to interpret neutron star cooling data without more information on the
value of $\eta$ at large densities, it also means that more neutron
star cooling data is essential to understanding the nature of the
high-density equation of state. Observations of neutron stars masses
and radii will have difficulty constraining the value of $\eta$.

Ref.~\cite{Gusakov05} studied in detail the cooling of neutron stars
constructed with the APR EOS. They found that, because of the direct
Urca process, stars with masses larger than about 1.7 $\mathrm{M}_{\odot}$ cool
sufficiently rapidly as to be cooler than nearly all of the observed
neutron stars. As these authors point out, this is somewhat sensitive
to the assumptions about the pairing interaction, and sufficiently
cool neutron stars with large mass may be difficult to
obeserve. Nevertheless, this work offers another possible
interpretation: the actual value of $\eta_{\mathrm{pot}}$ may be
sufficiently small at high density to turn off the direct Urca process
thus making the computed cooling curves match the comparatively warm
neutron stars which are given in the experimental data.

Proto-neutron star cooling (during the first minute when the core is
still hot) always contains a significant contribution from direct Urca
(e.g. see the review in Ref.~\cite{Prakash97}) since the finite
temperature allows momentum conservation to be easily
fulfilled. However, the proton fraction of the hot neutron star matter
dictates the flavor content of the associated neutrino signal and this
will be modified by the presence of quartic terms.

Ref.~\cite{Steiner05} pointed out that there is a correlation between
the threshold for direct Urca and the neutron skin thickness of
lead. This work clearly indicates that the correlation is predicated
on the implicit assumption that $\eta_{\mathrm{pot}}$ is nearly unity
for the relevant densities. This will also, for some models, impact
the suggested correlation between the direct Urca process and neutron
star radii~\cite{Horowitz02}, since quartic terms are important for
the former but not for the latter.

It might be interesting to explore the effect of a quartic term on the
symmetry energy at finite temperature. The value of $\eta$ for the
kinetic part of the equation of state could become closer to unity
since finite-temperature effects are proportional to
$T^2$. Nevertheless, various experiments are now probing the
temperature-dependence of the symmetry energy and might provide
constraints on a quartic behavior~\cite{Tsang01,Ono03,Kowalski06}.

The effects of a phase transition to hyperons, Bose-condensates
(except for that already present in APR), or quarks have been
ignored. It might be interesting to explore the value of $\eta$ for
equations of state involving phase transitions. Although quark matter
is often more isospin-symmetric, it also frequently has a significant
gap, which prohibits the direct Urca process for low enough
temperatures.

\section{Acknowledgements}

The author would like to thank Bao-An Li, Peter M\"{o}ller, Madappa
Prakash, and Sanjay Reddy for their helpful comments. This work was
carried out under the auspices of the National Nuclear Security
Administration of the U.S. Department of Energy at Los Alamos National
Laboratory under Contract No. DE-AC52-06NA25396.

\bibliography{sym4}
\bibliographystyle{apsrev}

\end{document}